\documentclass[conference,letterpaper]{IEEEtran}
\IEEEoverridecommandlockouts
\usepackage{cite}
\usepackage{amsmath,amssymb,amsfonts}
\usepackage{algorithm}
\usepackage{algorithmic}
\usepackage{graphicx}
\usepackage{textcomp}
\usepackage{xcolor}
\usepackage{url}
\usepackage{amssymb, graphicx, amsmath, amsthm}
\usepackage{tikz}
\usepackage{multicol, blindtext}
\usepackage{subfigure}

\def\BibTeX{{\rm B\kern-.05em{\sc i\kern-.025em b}\kern-.08em
    T\kern-.1667em\lower.7ex\hbox{E}\kern-.125emX}}
\begin{document}

\title{Using Geographic Load Shifting \\ to Reduce Carbon Emissions}

\author{
\IEEEauthorblockN{Julia Lindberg, Bernard C. Lesieutre and Line A. Roald}
\IEEEauthorblockA{
University of Wisconsin-Madison,
United States, 
Contact: jrlindberg@wisc.edu
}
}

\maketitle

\begin{abstract}
An increasing focus on the electricity use and carbon emissions associated with computing has lead to pledges by major cloud computing companies to lower their carbon footprint.
Data centers have a
unique ability to shift computing load between different geographical locations, giving rise to geographic load flexibility that can be employed to reduce carbon emissions.
In this paper, we present a model where data centers shift load independently of the ISOs. We first consider the impact of load shifting guided by locational marginal carbon emissions, denoted by $\lambda_{\text{CO}_2}$, a sensitivity metric that measures the impact of incremental load shifts. Relative to previous models for data center load shifting, the presented model improves accuracy and include more realistic assumptions regarding the operation of both data centers and the electricity market.
Further, we introduce a new benchmark model in which data centers have access to the full information about the power system and can identify optimal shifts for the current time period.
We demonstrate the efficacy of our model on the IEEE RTS GMLC system using 5 minute load and generation data for an entire year. Our results show that the proposed accuracy improvements for the shifting model based on $\lambda_{\text{CO}_2}$ are highly effective, leading to results that outperform the benchmark model. 
\end{abstract}


\section{Introduction}

The recent technology revolution has infiltrated most aspects of modern day life,  
and has lead to an increase in demand for computing resources. Between 2010 and 2018 there was an estimated 550\% increase globally in the number of data center workloads and computing instances \cite{masanet2020recalibrating}. This increased computing demand has commanded a  shift from smaller data centers to large-scale, highly optimized and efficient facilities. These facilities are referred to as \emph{hyper-scalar} data centers and are operated by large technology companies such as Amazon, Facebook, Google, Microsoft and Alibaba. These companies operate vast networks of these data centers which are dispersed geographically throughout the world \cite{nrdc_2014, yevgeniy_analysts}. 
Data centers currently consume 
around 1-2\% of electricity 

globally \cite{masanet2020recalibrating}. This low percentage is largely due to the increased efficiency of hyper-scalar data centers, which have processed the increased computing demand while only increasing electricity use by an estimated 6\% \cite{masanet2020recalibrating}. However, 
as the scope for further efficiency gains is 
largely exhausted, it is projected that electricity use for computing will increase rapidly in the future.

Hyper-scalar data centers are large loads on electric power networks with unique characteristics. 
For example, data center loads can defer when computing tasks are processed or process them at different locations. These properties equip data center operators with the unique ability to participate in both geographic and temporal load shifting. 
Motivated by recent pledges made by technology companies to reduce their carbon emissions \cite{googleenvironmentalreport, amazonenvironment,googleblog}, we are interested in understanding how these hyper-scale data centers can effectively interact with electricity markets in a way that minimizes the carbon emissions caused by their computing loads. Start-ups such as Lancium consider computing that adapts to the operation of the electric grid \cite{lancium-press}, an idea that plays a critical role in the vision of zero carbon cloud computing \cite{yang2016zccloud, chien2019zero}.

When shifting load, data centers interact with electricity markets operated by independent system operators (ISOs).
Previous research has examined the impact of integrating data centers and demand response \cite{dcdr_survey,liu2013data,liu2014pricing, zhou2016bilateral, zhou2020a, zhou2015when, chen2019an} or have considered geographical load shifting to reduce electricity costs \cite{ li_bao_li_2015, rao_liu_ilic_liu_2012, dou2017carbon}, including load shifting between different electricity markets \cite{rao_liu_xie_liu_2010}. Other work has  modelled data center flexibility through the use of virtual links in time and space \cite{zhang2019flexibility}. 
Much of this body of work considers the shifting of computing load to reduce the carbon emissions of data centers \cite{goiri2013parasol, liu2012renewable, dou2017carbon} and increase absorption of renewable energy \cite{kim_yang_zavala_chien_2017, zheng2020mitigating}.
There are, however, several major challenges to effective collaboration between the ISO and data centers, such as barriers to the exchange of possibly sensitive information. Another challenge is that ISOs and data centers have different objectives, namely, ISOs focus on minimizing electricity cost while data centers would like to minimize carbon emissions. 
While it is a common assumption that those two objectives are well aligned, 
our previous research \cite{lindberg2021a} found that if data centers are specifically targeting the objective of reducing carbon emissions (rather than reducing cost) they can achieve better results by 
doing the load shifting outside of the electricity market rather than collaborating with the ISO. Since ISOs are unlikely to change the market to directly minimize carbon emissions in the near future, we focus on a method to identify geographical load shifts that reduce carbon emissions within the existing electricity market.

To effectively shift load geographically, we require information about the locational variation in the carbon footprint of electricity, similar to the way locational marginal prices describe the locational variations in the cost of electricity.
While data on electricity prices is easily available, similar information on the carbon emissions associated with electricity usage is not. As a result, it is less straight forward to develop metrics that provide good guidance for shifting load to reduce carbon emissions. 
Previous work in this vein has assumed that prices are directly tied to the fraction of non-renewable energy \cite{2015liu}, or considered average carbon emissions for electricity in a region and/or renewable energy curtailment \cite{chien2015zero, yang2017large, zheng2020mitigating}. A benefit to these metrics is that several companies provide the necessary information needed to compute them \cite{tomorrow, caisoemissions,caisocurtailment}. A downside is that they all fail to consider necessary aspects of electric grid operation, such as the impact of a marginal change in load and transmission capacity limits. Specifically, there is no consideration for which generators in the system will be providing the additional generation or if is there is sufficient transmission capacity available to deliver the generation. 
To effectively guide load shifting, we need 
a measure of marginal emissions \cite{lindberg2021a}.
The use of marginal emissions to assess energy efficiency or the potential impact of renewable energy has been considered in \cite{siler2012marginal, callaway2018location}, proposed to 
guide renewable energy investments \cite{2010Ruiz, 2011Rudkevich, siler2012marginal}, 
or considered the impact of marginal emissions rates at cogeneration facilities \cite{tabors2021methodology}.
More recently, a measure of locational marginal carbon emissions was proposed for data center geographical load shifting in \cite{lindberg2021the}. The superiority of this metric over others such as average carbon emissions, curtailment or LMPs for data center geographic load shifting was demonstrated in \cite{lindberg2021a}.

While the previous load shifting models in \cite{lindberg2021the, lindberg2021a} presented promising results, it included several unrealistic assumptions such as the electricity market being solved twice in each time step. Here, we present a model that addresses these issues and also leverages regularization to increase the accuracy of the model. 
Further, due to the fact that the shifting metric derived in \cite{lindberg2021the} is a linear sensitivity, we expect that at each time step this metric is not necessarily finding the best load shift. For this reason, we present a new benchmark model for optimal load shifting. This bilevel program finds the best load shift a data center can perform, given the current electricity market.
In summary, the three main contributions of this paper are the following:
    \noindent
    \emph{1) Realistic model for data center load shifting:} We address the unrealistic assumptions in our previous model by implementing load shifting in a cumulative fashion and incorporating a regularization parameter to increase the accuracy of shifting.
    
    \noindent
    \emph{2) Benchmark model for optimal data center shifting:} We develop a new benchmark model that computes the optimal data center load shift, given access to all system information. This bilevel optimization method determines the optimal load shift to reduce carbon emissions, subject to the operation of the current electricity market.
    
    \noindent
    \emph{3) Computational Analysis:} We compare both models using the RTS-GMLC system with one year of 5 minute load and generation data, and observe that the shifting model based on locational marginal carbon emissions performs quite well.

The remainder of the paper is organized as follows.
In Section~\ref{sec:2} we review the existing data center driven shifting model, while Section~\ref{sec:3} presents the proposed improvements to this model. In Section~\ref{sec:4}, we present a new benchmark model for optimal load shifting. Section~\ref{sec:5} demonstrates the efficacy of our model in a case study, and Section \ref{sec:6}  concludes.

\section{Review of Existing Load Shifting Model}\label{sec:2}
Previous work has considered ISO independent load shifting to reduce carbon emissions via a value called the \textit{locational marginal carbon emissions}, $\lambda_{\text{CO}_2}$ \cite{lindberg2021the}. This value is calculated as a linear sensitivity around the optimal solution to the DC OPF. We give a brief outline of this load shifting model here, but suggest \cite{lindberg2021the} as a more detailed reference. 

\noindent \textbf{Step 1:} This model begins by assuming the ISO solves a DC OPF \cite{christie2000transmission,litvinov2010}. The DC OPF is a linear optimization problem that seeks to minimize generation costs subject to network and demand constraints with decision variables $x = [\theta \  P_g]$ where $P_g$ are the generation variables, $\theta$ are the voltage angles at each node. Let $\mathcal{N}$ be the set of all nodes, $\mathcal{L}$ the set of lines and $\mathcal{G}$ the set of generators. The DC OPF is defined as
\begin{subequations}\label{dcopf}
\begin{align}
\min_{\theta, P_g} ~~&c^T P_g \label{dcopfcost} \\
\text{s.t.} ~~& \textstyle \sum_{\ell\in\mathcal{G}_i}  \!P_{g,\ell} -\! \textstyle \sum_{\ell\in\mathcal{D}_i} P_{d,\ell} = &&
\nonumber\\
&\qquad\quad\textstyle\sum_{j:(i,j)\in\mathcal{L}} \!\!\!\!-\beta_{ij}(\theta_i \!- \!\theta_j), &&\forall i\in\mathcal{N} \label{balance}\\
-&P^{lim}_{ij} \!\leq\! -\beta_{ij}(\theta_i \!-\!\theta_j) \!\leq\! P^{lim}_{ij}, &&\forall (i,j)\in\mathcal{L}
\label{lineineq}\\
& P^{min}_{g,i} \leq P_{g,i} \leq P^{max}_{g,i}, && \forall i\in\mathcal{G}
\label{genineq}\\
& \theta_{ref} = 0. \label{refnode} 
\end{align}
\end{subequations}

The cost function \eqref{dcopfcost} minimizes the cost of generation, where $c_i$ is the cost of generation at generator $i$. Constraints \eqref{balance}-\eqref{genineq} constrain the nodal power balance, transmission line and generator capacity constraints respectively. Finally, \eqref{refnode} sets the voltage angle at the reference node to zero.

\noindent \textbf{Step 2:} Independent of any ISO collaboration, data center operators shift their load to minimize carbon emissions. To guide this effort, a metric known as the \textit{locational marginal carbon emissions} was proposed in \cite{lindberg2021the}. In \cite{lindberg2021a} the authors demonstrated the superiority of this metric to other more commonly studied metrics such as the average carbon emissions or excess low carbon power.  

To calculate the locational marginal carbon emissions, we first consider a basic optimal solution $x^* = [\theta^*, P_g^*] \in \mathbb{R}^n$ to the DC OPF \eqref{dcopf}. From sensitivity analysis in linear optimization theory \cite{bertsimas1997introduction}, a basic optimal solution can be written as $Ax^* = b$, where $A \in \mathbb{R}^{n \times n}$ is a full rank matrix consisting of all the active constraints of \eqref{dcopf} at the optimal solution $x^*$. In the case of the DC OPF, the rows of $A$ consist of the equality constraints \eqref{balance} and \eqref{refnode} as well as a subset of the inequality constraints \eqref{lineineq}, \eqref{genineq} that are satisfied at equality for $x^*$.

A small change in load can be represented as a small change in the right hand side $b$, given by $\Delta b = \begin{bmatrix} \Delta P_d & 0 \end{bmatrix}^T$. Assuming that the change is small enough as not to alter the set of active constraints, we compute the optimal change in generation as $A \cdot \Delta x = \Delta b$ where $\Delta x = [\Delta \theta \ \Delta P_g]$. This gives rise to the linear relationship
\vspace{-1mm}
\begin{align}
    \begin{bmatrix} \Delta \theta \\ \Delta P_g   \end{bmatrix} &= A^{-1} \cdot \begin{bmatrix} \Delta P_{d} \\ 0 \end{bmatrix}
\end{align}
If we denote the matrix consisting of the last $|\mathcal{G}|$ rows and first $|\mathcal{N}|$ columns of $A^{-1}$ by $B$, this gives the linear relationship between load and generation changes, $\Delta P_g = B \cdot  \Delta P_d$. 

Consider the cost vector $g \in \mathbb{R}^{| \mathcal{G}|}$ that measures the carbon emissions of each generator per MW. Specifically, the $i$th component of $g$, is the carbon intensity of generator $i$. Multiplying each side of $\Delta P_g = B \cdot  \Delta P_d$ on the left by $g$ gives us the change in carbon emissions
\vspace{-1mm}
\begin{align}
    \Delta CO_2 = g \cdot \Delta P_g = g \cdot B \cdot \Delta P_d = \lambda_{\text{CO}_2}  \Delta P_d \label{newobj}.
\end{align}
where $\lambda_{\text{CO}_2} = g \cdot B$. Intuitively, the $k$th component of $\lambda_{\text{CO}_2}$ measures how an increase of $1$ MW of load at node $k$ will affect the total carbon emissions of the system.
 
We let $\mathcal{C}$ denote the set of data center loads that can geographically shift load and consider optimization variables that denote the change in load at data center $i$, $\Delta P_{d,i}$, and the shift in load from data center $i$ to $j$, $s_{ij}$. The geographic load shifting optimization problem is given by:
\begin{subequations}\label{datacenteropt}
\begin{align}
\min_{\Delta P_{d}, s} \ \  & \sum_{i \in \mathcal{C}} \lambda_{\text{CO}_2,i} \Delta P_{d,i}  \label{loadshiftobj} \\
 \text{s.t. } \ \   &\Delta P_{d,i} = \textstyle\sum_{j\in\mathcal{C}} s_{ji} - \textstyle\sum_{k\in\mathcal{C}} s_{ik} \quad &&\forall i\in{C} \label{datacenterflex1} \\
    \textstyle &\sum_{i\in\mathcal{C}} \Delta P_{d,i}= 0 \label{datacenterflex2} \\
    &- \epsilon_i \cdot \text{Cap}_i \leq \Delta P_{d,i} \leq \epsilon_i \cdot \text{Cap}_i \quad \!\!&&\forall i\in\mathcal{C} \label{datacenterlim1} \\
    &0 \leq \Delta P_{d,i} + P_{d,i} \leq \text{Cap}_i \quad &&\forall i\in\mathcal{C} \label{datacenterlim3} \\
   & 0 \leq s_{ij} \leq M_{ij} \quad &&\forall ij\in\mathcal{C}\!\times\!\mathcal{C}. \label{datacenterlim2} 
\end{align}{}
\end{subequations}
The objective value \eqref{loadshiftobj}  minimizes the change in carbon emissions as a function of the change in load, \eqref{datacenterflex1} enforces that the change in load at a given data center is equal to the total load shifted in minus the total load shifted out, while \eqref{datacenterflex2} says the sum of all load shifts must be zero. Constraint \eqref{datacenterlim1} limits the amount each data center can shift as a fraction $\epsilon$ of the data center capacity $\text{Cap}$. Constraint \eqref{datacenterlim3} puts an upper bound $M_{ij}$ on the total capacity at each data center and \eqref{datacenterlim2} limits how much load data center $i$ can send to data center $j$.

\noindent \textbf{Step 3:} Finally, the ISO resolves the DC OPF \eqref{dcopf} with new load profile, $P_{d,i}' = P_{d,i} + \Delta P_{d,i}^*$, where $\Delta P_{d,i}^*$ is the optimal solution to \eqref{datacenteropt} for all $i \in \mathcal{N}$.

\section{Realistic Data Center Load Shifting  Model}\label{sec:3}

The above model has several drawbacks. First, it is unrealistic to assume that the ISO resolves the market clearing twice, once before and once after the shifting has happened. Second, since the model is linear, we tend to see large load shifts even with small differences in $\lambda_{\text{CO}_2}$ between data center locations. Since $\lambda_{\text{CO}_2}$ is a local sensitivity factor that is only accurate near the previous optimal solution, these large shifts lead to inaccurate results that sometimes increase carbon emissions. To address these issues, we introduce two improvements to the model: cumulative load shifting and regularization.

\subsection{Cumulative load shifts}

We refine the model defined in \cite{lindberg2021the} by considering \emph{cumulative load shifts}.
Instead of resolving the DC OPF in Step $3$ of the above model, the load shift is applied to the market clearing in the next time step. 
Specifically, the algorithm runs as follows:

\noindent\textbf{Step 1}: At time $t$, the ISO solves the DC OPF \eqref{dcopf} with data center load set to $P_d^{t}$.

\noindent\textbf{Step 2}: Given information about $\lambda_{\text{CO}_2}$ as described above, the data center operator computes a load shift $\Delta P_d^{t}$ according to \eqref{datacenteropt}. Then, the data center load for time $t+1$ is set to 
$P_d^{t+1}= P_d^{t} + \Delta P_d^{t}$, and the algorithm proceeds to Step $1$ of the next time step.

While the cumulative load shifting model more accurately reflects the current market set up, it introduces an additional inaccuracy in our model. The locational marginal carbon emission value $\lambda_{\text{CO}_2}$ at each data center is derived as a linearization from the operating point at time $t$, but the internal data center shifting optimization will only affect the market clearing at time $t+1$. The expectation is that since operating conditions remain similar between time steps, shifting with respect to $\lambda_{\text{CO}_2}$ will still lead to a decrease in total system carbon emissions.
We also note that cumulative load shifting can increase accuracy relative to the existing model, particularly the load shift allowed in each time step is small (i.e., only a small fraction $\epsilon$ can be shifted). In this case, changes in the data center load build up slowly over time. This is in contrast to our previous model, where the data center load was reset to the original value $P_d$ in each time step.

\subsection{Regularizing load shifts}
To discourage large load shifts which can cause oscillations and increased emissions,
we propose to use a regularization term (i.e., a quadratic penalty) that discourages large shifts. 
Specifically, this model replaces the objective value \eqref{loadshiftobj} with 
\[
\sum_{i \in \mathcal{C}} \lambda_{\text{CO}_2, i} \Delta P_{d,i} + \gamma \| \Delta P_{d,i} \|_2^2
\]
where $\gamma \in \mathbb{R}$ is a regularization parameter. The goal in using this regularization term is to discourage large shifts that lead to an increase in carbon emissions as well as increase the accuracy of the data center driven shifting model. However, the regularization term can also be interpreted as a quadratic cost on load shifting. This ensures that while small shifts are cheap and frequent, we only shift a large amount of load when there will be a large reduction in carbon emissions. 

Throughout the rest of this paper we refer to the model outlined in 
this section
as ($\lambda_{\text{CO}_2}-$shift).

\section{Benchmark model for optimal shifting}\label{sec:4}
Our next contribution is to introduce a new model to benchmark the data center driven shifting model. Since the shifts provided by $\lambda_{\text{CO}_2}$ are calculated by a linear sensitivity, they can be inaccurate, even giving shifting profiles that increase carbon emissions. 

The problem of identifying the optimal load shift data center operators should employ to minimize carbon emissions can be modelled as a bilevel linear program. 
The upper level problem identifies the optimal choice of load shift $\Delta P_d$ to minimize carbon, i.e., 
\begin{align}
    \min_{\Delta P_d,s, P_g^*} &\ g^T P_g^* 
    \nonumber \\
    \text{s.t. }~~&P_g^* = \arg \min \eqref{opt:dcopf_with_flex} \label{opt:bilevel} \tag{Opt-shift}\\
    &(\Delta P_d, s) \in \mathcal{P} \nonumber
\end{align}
Here, the last constraint represents the set of feasible load shifts from the data center perspective, i.e.,  $\mathcal{P}$ is the polytope of permissible load shifts defined by the constraints in \eqref{datacenteropt}..
The first constraint states that the generation values $P_g^*$ is the solution to the lower level optimization problem \eqref{opt:dcopf_with_flex}. This problem is a version of the standard DC OPF \eqref{dcopf} where the nodal balance constraints include the change in demand. Formally we write it as
\begin{align*} 
    &\min_{P_g, \theta} \ c^T P_g  \quad 
     \text{subject to} \\
     & \text{Constraints } \eqref{lineineq},\eqref{genineq},\eqref{refnode}  \tag{DC-shift} \label{opt:dcopf_with_flex} \\
     &  \sum_{\ell\in\mathcal{G}_i} \! P_{g,\ell} -\!\!\! \sum_{\ell\in\mathcal{D}_i}  \!(P_{d,\ell} + \Delta P_{d,\ell}) = \!\!\!\!\!\!\! \sum_{j:(i,j)\in\mathcal{L}} \!\!\!\!\!\!\!\!-\beta_{ij}(\theta_i  \!- \! \theta_j), && \!\!\!\forall i\!\in\!\mathcal{N} 
\end{align*}

As in $(\lambda_{\text{CO}_2}$-shift), we also consider cumulative load shifting in \eqref{opt:bilevel}. 
Specifically, at each data center $\ell \in \mathcal{C}$, at time step $t$ we assume the load 
$P_{d,\ell}$ 
in \eqref{opt:dcopf_with_flex} reflects the sum of new load $P_{d,\ell}$ from time $t$ and the load shift $\Delta P_{d, \ell}$ from time $t-1$. Herein out when we refer to the model \eqref{opt:bilevel} we assume it is employed with this cumulative load shifting.

\section{Computational Results}\label{sec:5}
We next analyze the efficacy in carbon reduction of $(\lambda_{\text{CO}_2}$-shift) versus \eqref{opt:bilevel}. 

\subsection{Accuracy of $(\lambda_{\text{CO}_2}$-shift)}
An important aspect of $(\lambda_{\text{CO}_2}$-shift) that needs to be considered is the accuracy of the predicted change to carbon emissions and generation cost. Specifically, using the relation 
$\Delta P_g = B \cdot \Delta P_d $,
we can get a predicted change in generation that will result as an effect of the load shift $\Delta P_d$. Using this predicted generation change, we can derive a value for the predicted change in cost and carbon emissions to the system as a result of the load shift. 
We compare this predicted change to the true change in carbon emissions.

\subsection{Test Case}
We perform an extensive year long analysis of carbon reduction methods mentioned above using the RTS-GMLC system \cite{barrows2020the}. This system has $73$ buses, $158$ generators and $120$ lines. Since the original system does not designate data center loads, we assign data centers at buses $103, 107, 204$ and $322$. We assume that cumulatively, the four data centers consume a fixed power of $1000$ MW at each time step throughout the year, although the distribution of that power among the four data centers varies. We assume at time step $0$, each data center starts with $250$ MW of load. For all other loads and renewable generation, we use the real time, i.e. $5$ minute, load and generation data provided with \cite{barrows2020the}. Over the course of the year, this system serves $526,220,000$ MW of load, and $105,408,000$ MW or roughly $20.03\%$ of it is data center load.

Adding these large data center loads to the network greatly increases the total system load and results in time steps where the original DC OPF is infeasible. To remedy this we change the generation limits by setting $P_g^{\text{min}} = 0$ for all $g \in \mathcal{G}$, and increase the maximum generation limits by $50\%$. At each time step we allow each data center to shift up to $20\%$ of its total capacity, i.e. $50$ MW, and enforce that data center capacities remain between $0$ and $300$ MW. Further, we put no limitations on how much load each data center can shift to one another.

\subsection{The effect of regularization}
\begin{figure}
\begin{subfigure}
  \centering
  \includegraphics[width=.8\linewidth]{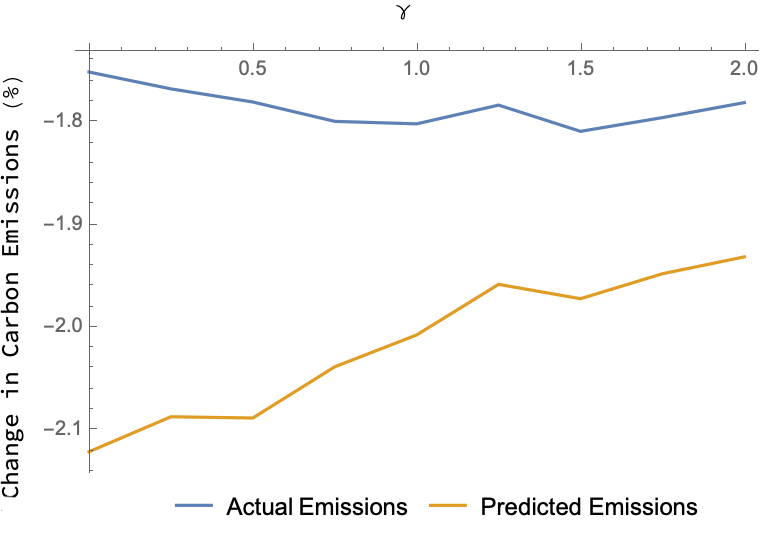}
  \caption{Change in carbon emissions.}
  \label{fig:sfig1}
\end{subfigure}%
\begin{subfigure}
  \centering
  \includegraphics[width=.8\linewidth]{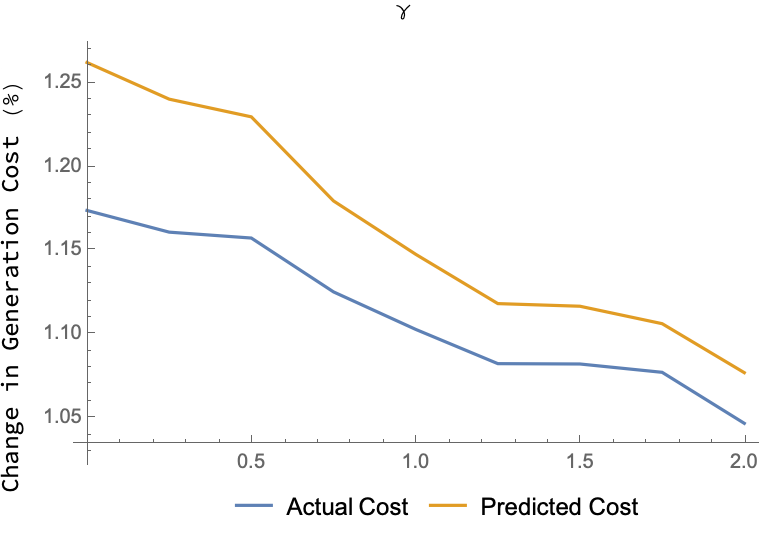}
  \caption{Change in generation cost.}
  \label{fig:sfig2}
\end{subfigure}
\begin{subfigure}
  \centering
  \includegraphics[width=.8\linewidth]{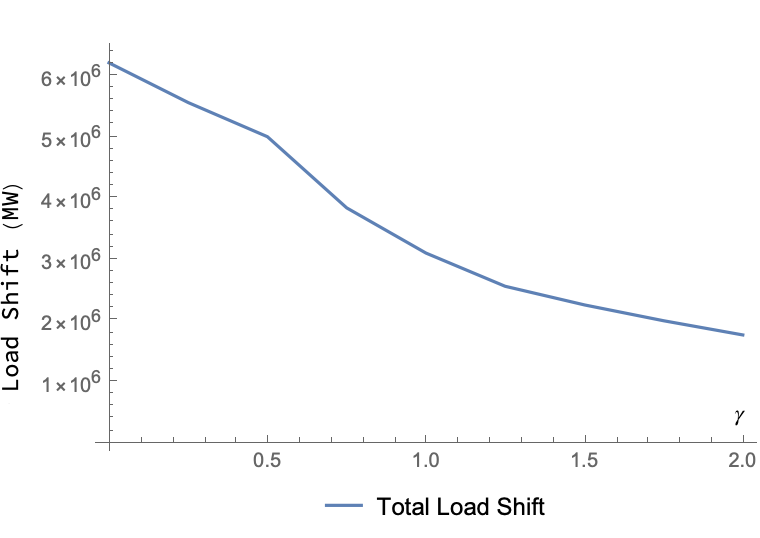}
  \caption{Change in total load shift.}
  \label{fig:sfig3}
\end{subfigure}
\caption{Change in carbon emissions, generation cost and load shift as the regularization parameter $\gamma$ varies.}
\label{fig:gamma_varies}
\end{figure}

We first investigate the effect of the regularization parameter $\gamma$. The effect of various regularization parameters on generation cost, total system carbon emissions and total load shift is shown in Figure~\ref{fig:gamma_varies}, where the orange and blue lines represent the predicted and actual values, respectively. Figure~\ref{fig:gamma_varies}(a) show that the minimum total system carbon emissions occurs when the regularization parameter $\gamma = 1.5$ is used. In addition, we see in both Figure~\ref{fig:gamma_varies}(a) and (b) that as the regularization parameter $\gamma$ increases, the difference between the predicted carbon emissions and generation cost and the actual carbon emissions and generation costs decreases. This indicates that including a regularization term helps not only in the efficacy of the data center driven shifting model, but also in the accuracy.

The reason regularization is considered is to discourage load shifting in cases where it is not predicted to make large differences. Figure~\ref{fig:gamma_varies}(c) shows how as the regularization parameter increases, the total load shifted throughout the year decreases. 
We see that when the regularization parameter is set at $\gamma = 1.5$, the total amount of load shifted is less than half of the amount of load shifted when $\gamma = 0$. Considering that the carbon emissions and generation cost when $\gamma = 1.5$ are lower than when $\gamma = 0$, this demonstrates that shifting less load, more strategically can lead to a larger reduction in carbon emissions and a smaller increase in generation costs.


\begin{table*}[h!]
    \centering
    \begin{tabular}{|c|c|c|c|c|}
    \hline 
           & DC OPF & \eqref{opt:bilevel}     & $\lambda_{\text{CO}_2}-$shift: $\gamma = 0$ & $\lambda_{\text{CO}_2}-$shift: $\gamma = 1.5$ \\
          \hline 
         Generation Cost& $3,802,706,000$ & $4,981,076,000$ &    $3,847,332,000$ & $3,843,847,000$ \\
         CO$_2$ Emissions & $164,402,000$ & $110,444,000$ &  $161,522,000$ & $161,427,000$ \\
         Total Shifts & 0 & $1,048,000$  & $6,199,000$ & $2,245,000$\\
         \hline 
    \end{tabular}
    \vspace{1mm}
    \caption{Summary of results from all models}
    \label{tab:result_summary}
\end{table*}

\subsection{Comparison with Opt-Shift and Original DC OPF solution}
We next compare the solutions for $(\lambda_{\text{CO}_2}$-shift) with regularization parameters $\gamma=0$ and $\gamma=1.5$ with the original DC OPF solution and the solution obtained using our benchmark model \eqref{opt:bilevel}.
These results are given in Table~\ref{tab:result_summary}. We see that when considering ($\lambda_{\text{CO}_2}$-shift) with no regularization, carbon emissions relative to the original DC OPF decreases by around 2.8 million tons or $1.75\%$. This reduction is achieved while shifting around $6.2$ million MW of load. Conversely, once the regularization term $\gamma=1.5$ is added, we achieve an even greater reduction in carbon emissions, namely $2,975,000$ tons or $1.81 \%$ while only shifting around $2.25$  million MW of load. In addition, when considering regularization, total system generation costs only increased by $1.08\%$ while without regularization it increased by $1.17 \%$.

In contrast to the above results, we see a dramatic carbon savings when using the benchmark \eqref{opt:bilevel}. In this case we save $53,958,000$ tons of carbon, i.e. $32.82 \%$. This occurs while only shifting a little over $1$ million MW. This dramatic savings occurs at a major increase to generation costs. Namely, \eqref{opt:bilevel} results in an increase in $\$ 1,178,370,000$ to generation costs or $30.99 \%$ over the original DC OPF. This benchmark model suggests that dramatic reductions in carbon emissions are possible even with limited data center flexibility, but come at a large increase to generation costs.

\subsection{Carbon Emissions vs Generation Costs}\label{ss:first_results}

\begin{figure}
    \centering
    \includegraphics[width = 0.4\textwidth]{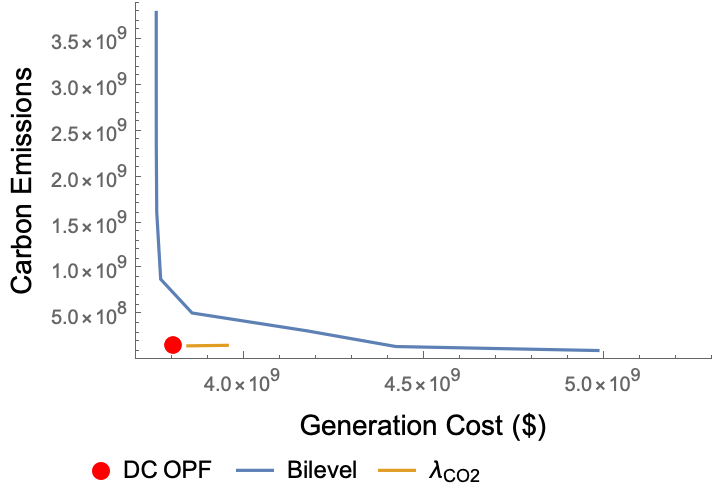}
    \caption{Trade off between carbon emissions and generation cost.}
    \label{fig:gen_constrained_bilevel}
\end{figure}

As seen above, 
minimizing carbon emissions can lead to an increase in generation cost.
To better understand the trade-off between carbon emissions and cost, we consider the benchmark model \eqref{opt:bilevel} with objective function
\[
(\alpha c^T + (1 - \alpha)g^T)P_g^*
\]
and $(\lambda_{\text{CO}_2}-$shift) with objective function
\[
(\alpha \text{LMP} + (1 - \alpha)\lambda_{\text{CO}_2})\Delta P_d + 1.5 \cdot \| \Delta P_d \|_2^2
\]
in place of \eqref{loadshiftobj} 
where $\alpha \in [0,1]$ is a trade off parameter that allows us to weight the emphasis on minimizing carbon emissions versus generation costs and LMP is a vector of the locational marginal prices at each node. 
The trade off between minimizing carbon emissions and generation cost is shown graphically in Figure~\ref{fig:gen_constrained_bilevel}.

When considering ($\lambda_{\text{CO}_2}-$shift), shown in yellow, we see a small variation in the overall system generation cost and carbon emissions that remains close to the carbon emissions and generation cost of the original DC OPF. This is consistent with the results shown above, and is due to the fact that this model considers small shifts away from an operating point that minimizes generation costs. 
The benchmark model \eqref{opt:bilevel} produces a much larger variation in operating points as we change the trade-off parameter $\alpha$. 
As $\alpha$ increases, the model produces a large increase in carbon emissions for only a moderate cost savings. In addition, we see that for \eqref{opt:bilevel} to achieve lower carbon emissions than the DC OPF and the $(\lambda_{\text{CO}_2}-$shift), a major increase to generation cost is needed. This demonstrates that even with limited geographic load shifting flexibility, a large reduction in carbon emissions is possible but it comes at the price of significantly higher generation costs.

Figure~\ref{fig:gen_constrained_bilevel} also demonstrates an interesting phenomenon, namely the greedy nature of \eqref{opt:bilevel}. 
When only trying to minimize carbon emissions, \eqref{opt:bilevel} is able to reduce total system carbon emissions by roughly $33 \%$ but this comes at a significant increase to total system generation cost. 
However, for the same generation cost, \eqref{opt:bilevel} gives a solution with higher carbon emissions than the DC OPF or $(\lambda_{\text{CO}_2}$-shift). This demonstrates that the greedy nature of \eqref{opt:bilevel} is not necessarily an optimal way to shift load over a long time span. Specifically, \eqref{opt:bilevel} finds a load shift that gives the largest reduction in carbon emissions at that time step, with no consideration to how the load shift will affect the carbon emissions of the system at the next time step. Using forecasts of future load and generation information to aid in a long term load shifting strategy is left as future work.

\subsection{Data Center Operating Load}
Finally, we show consider the impact of each model on the data center operating load. We consider $(\lambda_{\text{CO}_2}$-shift) with regularization parameter $\gamma = 1.5$ and \eqref{opt:bilevel}, and two different limits on the amount of load that can be shifted in each time step, $\epsilon = 0.01$ and $\epsilon = 0.2$. 

\begin{figure}[h!]
     \centering
     \includegraphics[width = 0.2\textwidth]{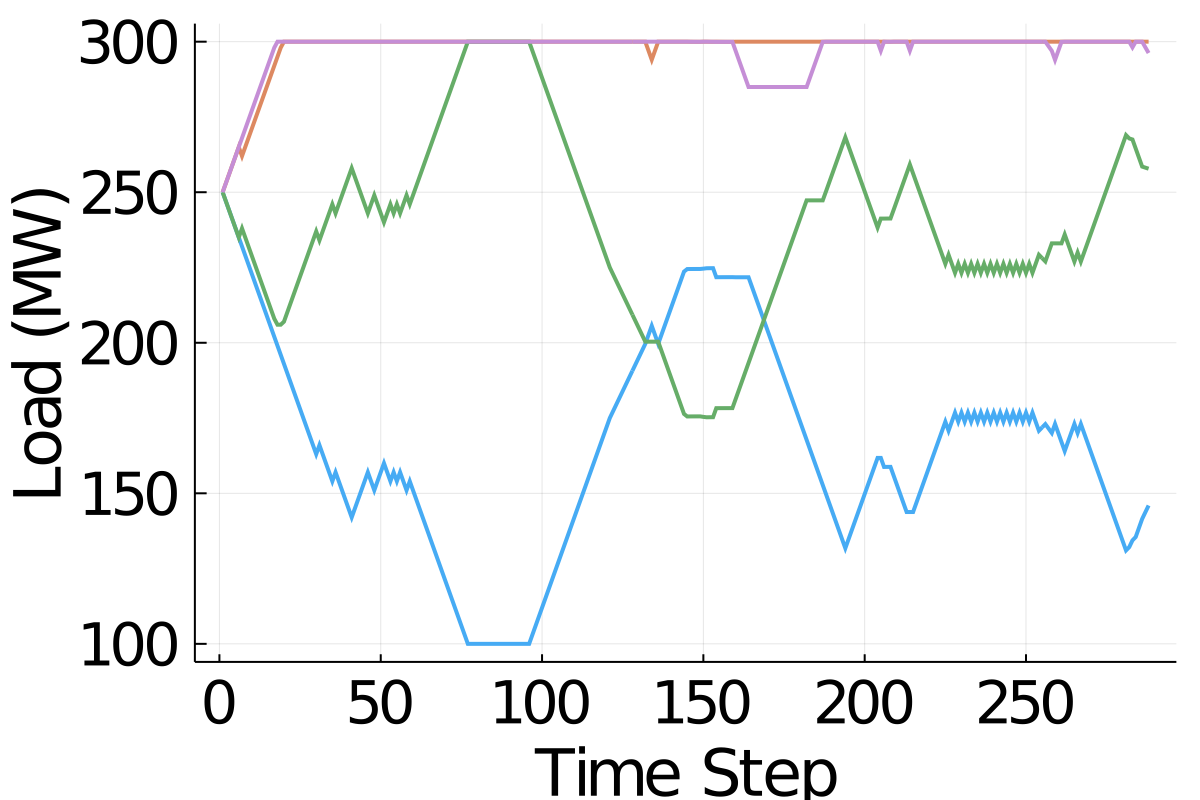}
     \includegraphics[width = 0.2\textwidth]{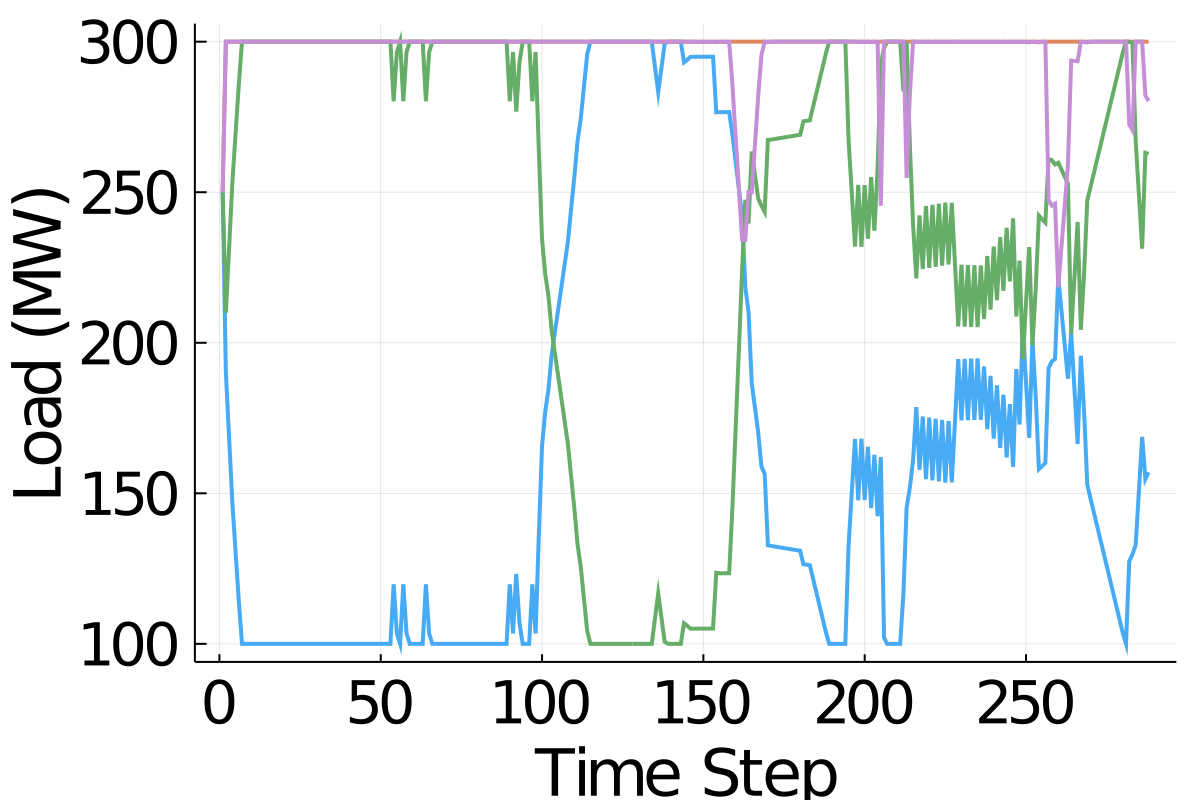}
     \includegraphics[width = 0.075\textwidth]{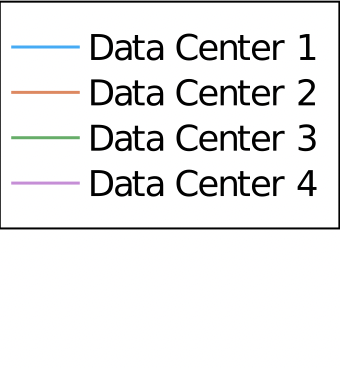}
     \caption{Load at each data center during the first $24$ hours using $\lambda_{\text{CO}_2}$-shift with $\gamma = 1.5$ and $\epsilon = 0.01$ (left) and $\epsilon = 0.2$ (right).}
     \label{fig:dc_load_lambda_CO2}
 \end{figure}

\begin{figure}[h!]
     \centering
     \includegraphics[width = 0.20\textwidth]{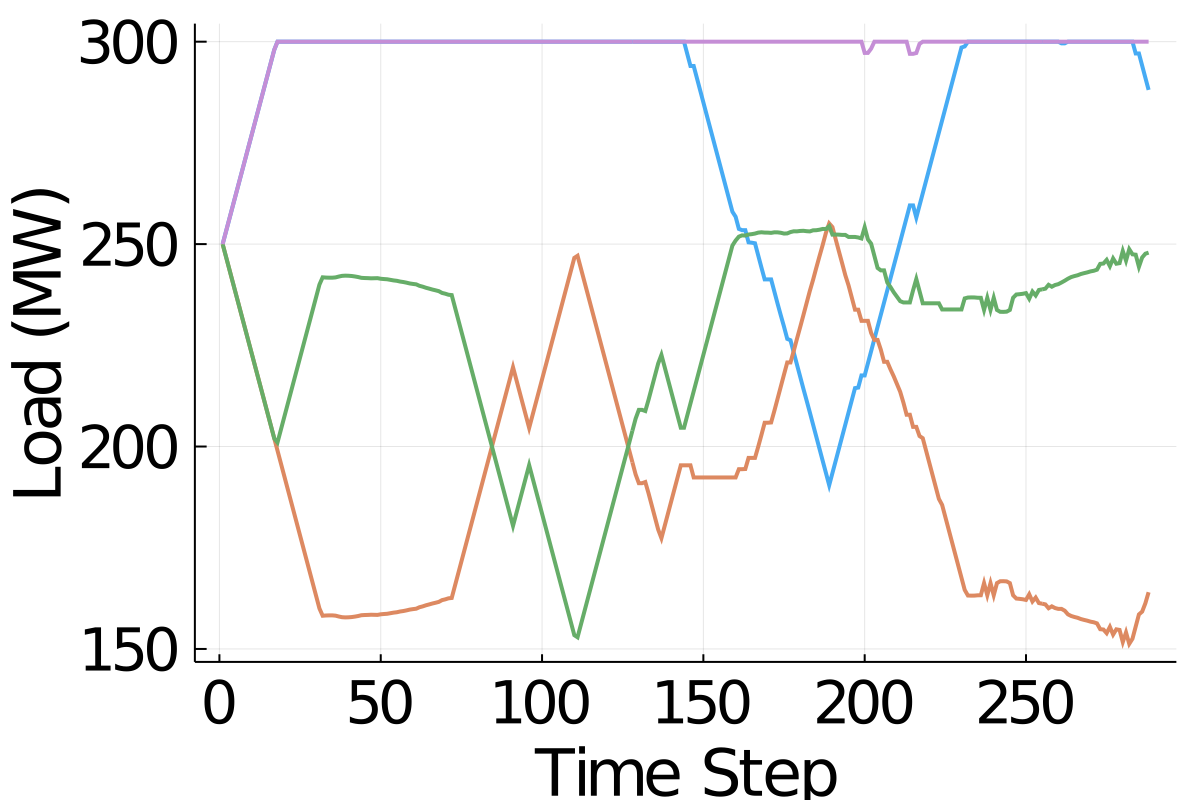}
     \includegraphics[width = 0.20\textwidth]{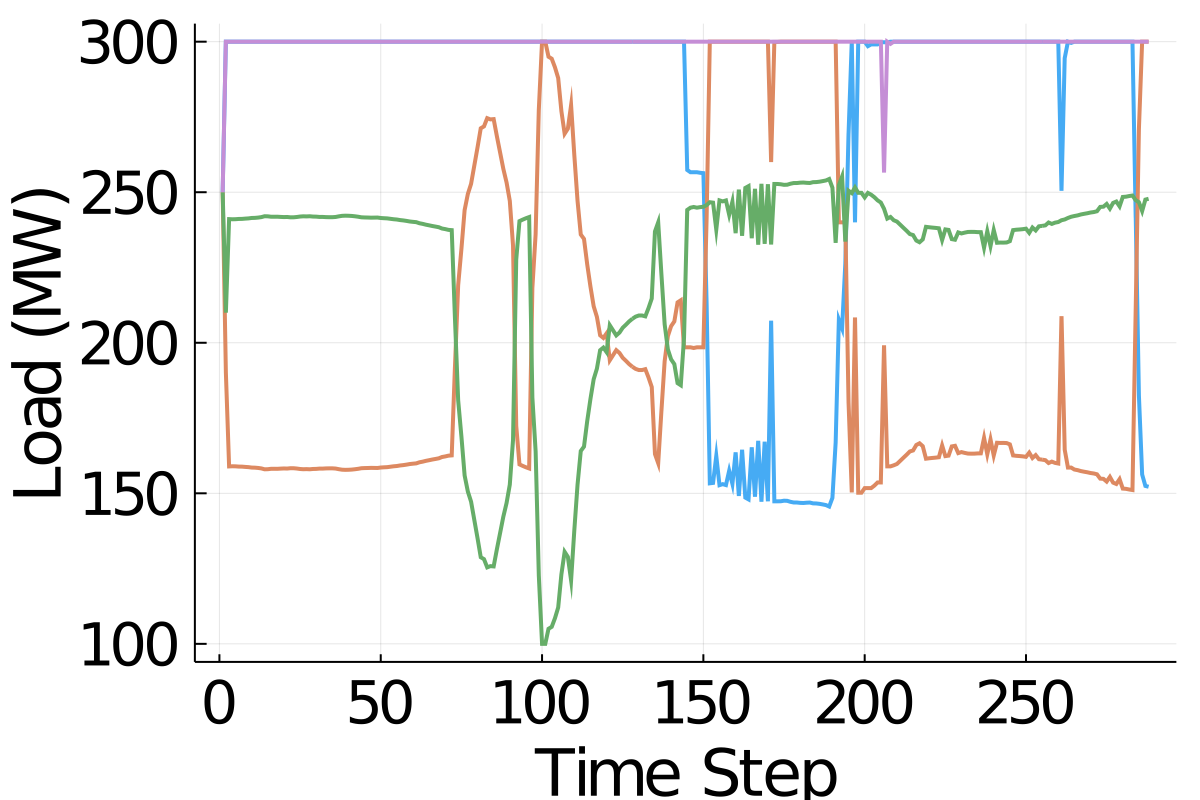}
     \includegraphics[width = 0.075\textwidth]{legend.png}
     \caption{Load at each data center during the first $24$ hours using \eqref{opt:bilevel} with $\epsilon = 0.01$ (left) and $\epsilon = 0.2$ (right).}
     \label{fig:bilevel_load}
 \end{figure}

In Figure~\ref{fig:dc_load_lambda_CO2} we see the operating conditions of each data center over the course of the first day when using $(\lambda_{\text{CO}_2}$-shift) when $\epsilon = 0.01$ (left) and $\epsilon = 0.2$ (right). In both cases we see similar overall trends in operating load.
However, $\epsilon = 0.2$ leads to much quicker changes and also dramatic oscillations in the load at data centers $1$ and $3$ towards the end of the day.
Similarly, in  Figure~\ref{fig:bilevel_load} we see the operating conditions of each data center over the course of the first day using \eqref{opt:bilevel} when $\epsilon = 0.01$ (left) and $\epsilon = 0.2$ (right). Again, we see similar trends data center load for both values of $\epsilon$, but for $(\lambda_{\text{CO}_2}$-shift), $\epsilon = 0.2$ leads to more oscillations in data center operating load.

Interestingly, there are some differences between the $(\lambda_{\text{CO}_2}$-shift) and  \eqref{opt:bilevel}. In both cases we see an initial pull for data center $4$ to operate at maximum capacity while the other data centers operate at lower capacities. This implies that the $\lambda_{\text{CO}_2}$ value for data center $4$ is accurately dictating that it is the most carbon neutral data center. In contrast, we see that when shifting with respect to $(\lambda_{\text{CO}_2}$-shift), data center $2$ is also operating at maximum capacity. This is in contrast to shifting when using \eqref{opt:bilevel}. In this instance data center $2$ initially drops to be the data center operating at the lowest load. This discrepancy highlights the inaccuracy when shifting with respect to $\lambda_{\text{CO}_2}$.

\begin{figure}[h!]
     \centering
     \includegraphics[width = 0.22 \textwidth]{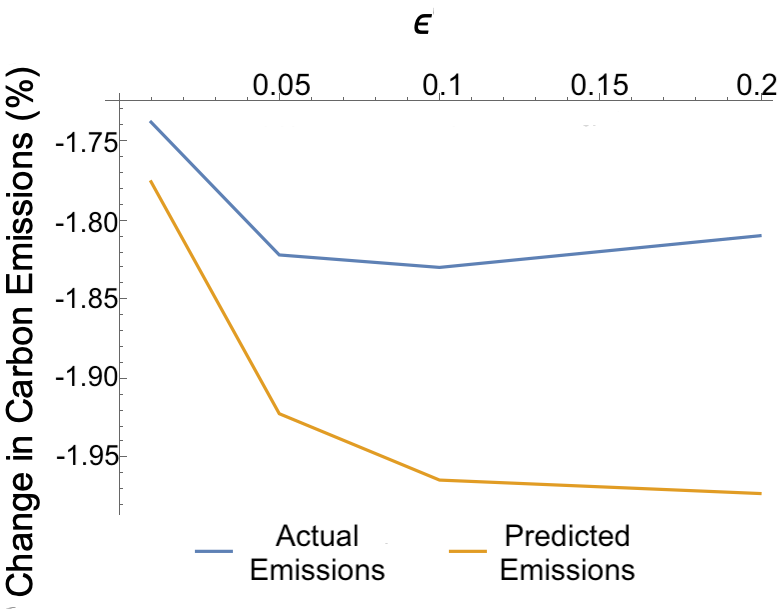}
     \includegraphics[width = 0.22 \textwidth]{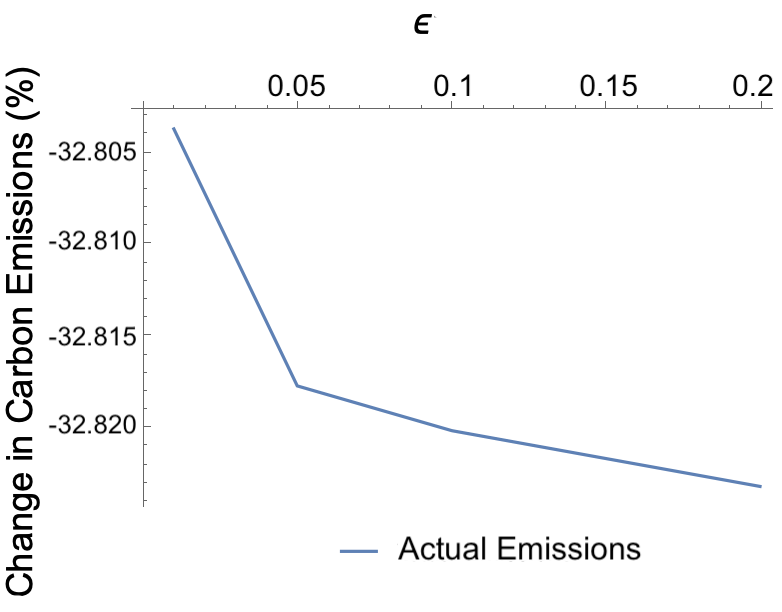}
     \caption{Predicted and actual change in carbon emissions using $\lambda_{\text{CO}_2}-$shift (left) and the change in carbon emissions using \eqref{opt:bilevel} (right) for varying epsilon values.}
     \label{fig:vary_epsilon}
 \end{figure}

 Finally, we investigate how using different $\epsilon$ values impacts the overall effect on carbon emissions.
 In Figure~\ref{fig:vary_epsilon} we see the change in total system carbon emissions as $\epsilon$ varies for both $(\lambda_{\text{CO}_2}$-shift) with $\gamma=1.5$ as well as \eqref{opt:bilevel}. In both cases we see only a very mild decrease in total carbon emissions as we allow $\epsilon$ to increase. Further, for $(\lambda_{\text{CO}_2}$-shift), we see that as $\epsilon$ increases, the accuracy of the model decreases and once $\epsilon >0.1$, the carbon emissions starts to increase. This indicates that allowing small shifts not only is more desirable from an operational stand point to avoid rapid changes and oscillations in data center loading, but it leads to similar carbon savings as allowing larger shifts.
 
\section{Conclusion}\label{sec:6}

In this paper we presented an improved model for data center load shifting to reduce carbon emissions. This model shifts load independently of ISO collaboration via a measure known as locational marginal carbon emissions. We built on existing work, but made several improvements to increase realism and accuracy of our model. 
We also proposed a new benchmark model which gives the best load shift at each time step, and compared the results.
The main conclusion from the paper is that smaller load shifts, limited by regularization and shifting caps, are quite effective in reducing carbon emissions. Larger load shifts tend to decrease accuracy of the model and produce less carbon savings. Further, while our benchmark model is able to achieve large carbon reductions, it also significantly increases cost. 

This paper demonstrates many natural directions for future work. First, this work shows that shifting load in a greedy way, i.e. shifting to reduce the maximal amount of carbon at each time step, is not necessarily the best approach if current load shifts will impact the load profile in future time steps.  
This demonstrates that forecasting future load and generation patterns is important to obtain better solutions. 
Finally, the information needed to calculate the locational marginal carbon emissions is currently not made publicly available. Therefore, finding ways to infer and predict $\lambda_{\text{CO}_2}$ at the data center nodes from publicly available data is needed in order to implement $(\lambda_{\text{CO}_2}$-shift) in practice.

\bibliographystyle{unsrt}
\bibliography{bibfile}

\end{document}